# Neutron Diffraction Studies on Chemical and Magnetic Structure of Multiferroic $PbFe_{0.67}W_{0.33}O_3$

Shidaling Matteppanavar[1], Basavaraj Angadi[1,*], Sudhindra Rayaprol[2]

[1]Department of Physics, JB Campus, Bangalore University, Bangalore –560056
[2]UGC-DAE-Consortium for Scientific Research, Mumbai Centre, BARC Campus, Mumbai – 400085
*Corresponding Author, e-mail: brangadi@gmail.com

**Abstract.** We report on the single phase synthesis and room temperature structural characterization of $PbFe_{0.67}W_{0.33}O_3$ (PFW) multiferroic. The PFW was synthesized by low temperature sintering, Columbite method. Analysis of powder XRD pattern exhibits single phase formation of PFW with no traces of pyrochlore phase. Detailed analysis of room temperature neutron diffraction (ND) reveals *cubic* phase at room temperature, space group *Pm-3m*. The ND pattern clearly reveals magnetic Bragg peak at $2\theta = 18.51^o$ ($Q = 1.36 Å^{-1}$). The refinement of magnetic structure reveals G-type antiferromagnetic structure in PFW at room temperature. The dielectric constant and loss tangent decreases with increasing frequency. The room temperature P-E measurements shows a non-linear slim hysteresis, typical nature of relaxor multiferroics, with saturation and remnant polarizations of $P_s = 1.50$ μC/cm$^2$ and $P_r = 0.40$ μC/cm$^2$, respectively.

## INTRODUCTION

Magnetoelectric (ME) multiferroics have been investigated extensively in recent years due to the coexistence of magnetic and electric ordering. They also exhibit great potential for the applications in multifunctional devices. Although a large number of ferroic materials (i.e. ferroelectric, ferromagnetic and ferroelastic) are available nowadays in different forms and crystal structure, the discovery of multiferroic phenomena in some materials has attracted much attention of scientists and technologists to develop new materials (or systems with optimum properties) for multifunctional devices [1-4]. Such devices can be fabricated using multiferroic materials, where a strong coupling between magnetic and electric order parameters exists in distorted structures.

Multiferroics have applications in memory devices, spintronics, microelectronics, magnetic sensors and electrically tunable microwave devices like filters, oscillators and phase shifters [1]. Among all these room temperature (RT) multiferroics discovered so far $PbFe_{0.67}W_{0.33}O_3$ (PFW) has unique properties with a high degree of order parameters and practically viable magnetic (paramagnetic-to-antiferromagnetic ordering at the $T_N$ ~ 350-380 K) and ferroelectric phase transition temperatures (paraelectric-to-ferroelectric phase transition 150-200 K) [2]. There have been some recent reports on the ME coupling between the ferroelectric and antiferromagnetic orders in PFW and related materials confirming the existence of the anomaly in dielectric constant [3], lattice parameters [2] at $T_N$ as well as a change in the dielectric constant induced by an external magnetic field [3].

In spite of the above qualities, this material has some inherent problems that limit its applications. It is well reported in the literature that synthesis of single phase PFW with perovskite phase has been difficult by the conventional methods, due to the formation of unwanted pyrochlore phases. Other alternative methods either involve two or more calcination steps or high sintering temperatures. For fabrication of materials on industrial scale, a simple and reliable method of synthesis is desirable.

In the present work, we have employed the Columbite method and sintering at low temperature to achieve single phase PFW. The sample has been characterized for the nuclear (crystallographic) and magnetic structure at room temperature using neutron diffraction of powder sample. We have also carried out dielectric measurements on the sample, to determine the ferroelectric properties.

## EXPERIMENTAL

PFW has been prepared by Columbite (solid-state reaction) method [5, 6] using high-purity (~99.9%) precursors such as PbO, $Fe_2O_3$ and $WO_3$ (reagent grade). $Fe_2O_3$ and $WO_3$ were taken stoichiometrically and ground in agate pestle and mortar in ethanol medium for 2 hours. The dried $Fe_2WO_6$ (Columbite) powder was calcined at 1000 °C 4 hours. After the 1st calcination stage the powder was ground again and stoichiometrically weighed PbO mixed with $Fe_2WO_6$. After the 2nd calcination (900 °C 2 hours) stage the PFW powder was ground again and polyvinyl alcohol (PVA) was added. Pellets of 10mm (5mm) in diameter and 2-3mm thickness were uniaxially pressed at 50 kN using a hydraulic press. The pellets were sintered at 850 °C for 90 min in a closed Pb rich environment to minimize the PbO evaporation. The phase formation of PFW samples were determined by X-ray powder diffraction using Phillips (1070 model) diffractometer with Cu-$K_\alpha$ ($\lambda$ = 1.5406 Å) radiation.

Neutron diffraction (ND) measurements were carried out on FCD (PD-3), a PSD-based powder diffractometer at Dhruva reactor BARC, using a wavelength of ($\lambda$ = 1.48 Å). Structural analysis was carried out by using Rietveld refinement of the ND data was using the *Fullprof* suite programs. For electrical characterization, both the planar surfaces of the sample were coated with high-purity silver paint. The hysteresis loop of the poled sample was traced from the P–E hysteresis loop tracer (sawer-tower circuit). The dielectric/impedance parameters, such as capacitance (C), dissipation factor and impedance (Z), were measured in a wide range of frequency (100 Hz–10 MHz) at room temperature (300 K) using a phase-sensitive LCR/impedance meter (model: PSM-1735).

## RESULTS AND DISCUSSION

The X-ray diffraction (XRD) pattern of sintered PFW recorded at room temperature is shown in Figure 1. All the peaks were indexed and are in good agreement with standard JCPDS pattern no. 32-0522 for PFW confirming the single phase formation. The formation of unwanted pyrochlore phases was completely bypassed in this method and this method is helpful in producing the 100% perovskite phase. The low temperature calcination along with the closed and Pb rich environment helps in minimizing the PbO evaporation, which has been known to be the origin for the pyrochlore phase formation. Heating in the temperature 850 – 1000°C for a total duration of about 8 hrs turned out to be the optimum condition for producing single phase PFW sample.

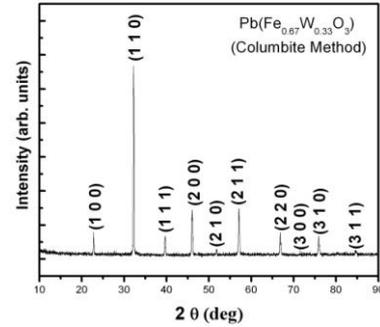

**FIGURE 1.** Room temperature X-ray diffraction pattern of PFW. The Miller indices have been indicated for each peak.

The room temperature neutron diffraction data of the sintered PFW is shown in Fig. 2. The structural analysis of the nuclear structure reveals the cubic structure of PFW space group, *Pm-3m*.

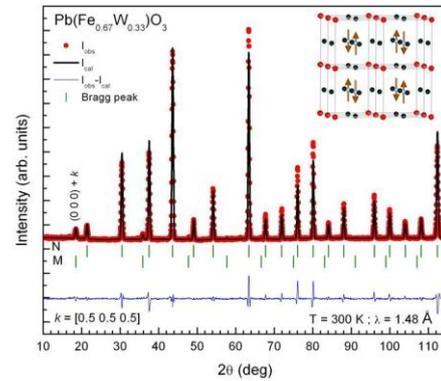

**FIGURE 2.** Rietveld refinement of neutron diffraction pattern of PFW at 300K. The structure (in 2a x 2b x 2c format) of PFW is shown as an inset. The arrows indicate the arrangement of spin on Fe atoms.

The R-factors for the refinement of ND data are $R_p$= 4.8; $R_{wp}$= 6.71; $R_{exp}$= 2.86; $\chi^2$ = 5.51. The refined cell parameter is *a* = 3.9828(4) Å. The ND pattern exhibits magnetic Bragg peaks at angles 2θ = 18.51, 35.95 and 47.81, confirming the magnetic state of the sample at room temperature. This also shows that the $T_N$ is above the room temperature. Magnetic structure was refined using a commensurate propagation vector, *k* = (½ ½ ½). The optimum agreement with diffraction data (expressed by the minimum value of $R_{mag}$) was obtained for an antiferromagnetic, G-type magnetic structure (inset of Fig. 2). Irreducible representations of little group and basis vectors were obtained using the *BasIreps* program available with in the *Fullprof* suite.

The P-E hysteresis loop for the PFW measured in a field of ±4 kV/cm is shown in Fig. 3. The room temperature non-linear slim hysteresis above the Curie temperature clearly indicates the typical relaxor nature and can be attributed to the existence of nanopolar clusters and the large electric field applied, which tends to align the nanopolar cluster into long-range order [7, 8]. The observed remnant polarization and coercive field values are at 50 Hz $P_R \approx 4$ μC/cm² and $E_C \approx 7$ kV/cm, respectively.

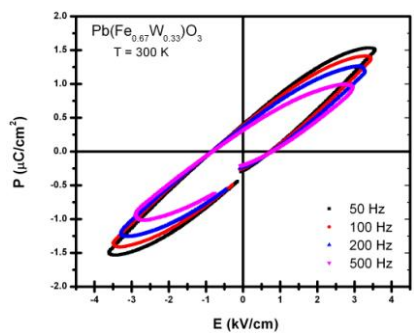

**FIGURE 3.** P-E Hysteresis loop for the sintered PFW at room temperature.

The P-E hysteresis loop for the PFW measured in a field of ±4 kV/cm is shown in Fig. 3. The room temperature non-linear slim hysteresis above the Curie temperature clearly indicates the typical relaxor nature and can be attributed to the presence of some microscopic polar regions. The observed remnant polarization and coercive field values are at 50 Hz $P_R \approx 4$ μC/cm² and $E_C \approx 7$ kV/cm, respectively.

Fig.4 (a) shows the variation of dielectric constant and *ac* conductivity verses frequency obtained for PFW at room temperature in the frequency range 100 Hz -10 MHz showing the decreasing trend in both real and imaginary parts of dielectric constant with increasing frequency. However the *ac* conductivity shows the reverse trend.

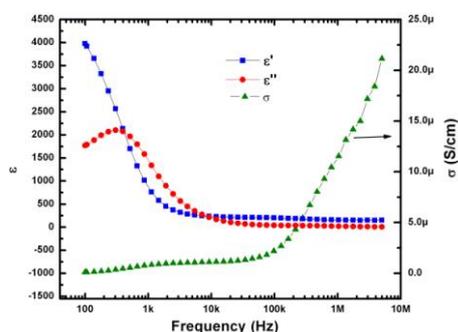

**FIGURE 4.** (a) Dielectric constant (ε′ and ε″) and ac conductivity (σ) vs. frequency for PFW at 300K

Fig.4 (b) shows the variation of the imaginary component of impedance with its real part (i.e. Z′ Vs Z″ plot). This type of plot is known as Nyquist plot or Cole–Cole plot. For a Debye-type relaxation, a single semicircle is observed (with its centre on Z′-axis), but Figure 4 (b) shows depressed semicircle, which suggests a deviation from the ideal Debye-type behavior [9, 10].

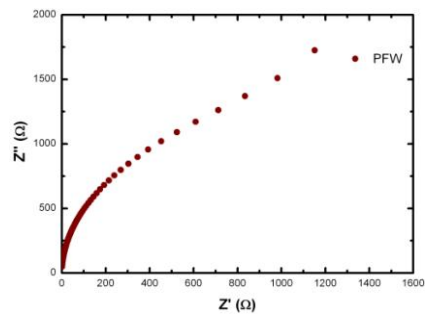

**FIGURE 4.** (b) Variation of Z′ with Z″ of PFW at 300K

In conclusions, the multiferroic PFW was successfully synthesized by Columbite method using lower calcination and sintering temperatures. Using the above mentioned synthesis conditions, PFW in single-phase, forms in perovskite structure with no traces of pyrochlore phase. The room temperature XRD and neutron diffraction studies showed that the synthesized PFW is in the cubic phase with space group *Pm-3m*. The magnetic structure of PFN refined using room temperature ND data reveled to be an antiferromagnetic with G-type magnetic structure. The room temperature dielectric measurement with frequency reveals the decreasing dielectric constant and increasing ac conductivity with frequency, a typical nature of dielectric material. The samples exhibit relaxor type P-E hysteresis with the obtained values of $P_R \approx 4$ μC/cm² and $E_C \approx 7$ kV/cm at 50 Hz.

## ACKNOWLEDGMENTS

The authors thank the UGC-DAE-CSR, Mumbai Centre for the financial support through CRS-M 159. The authors thank Dr. V. R. Reddy (UGC-DAE-CSR-Indore) and Dr. S. K. Deshpande (UGC-DAE-CSR-Mumbai) for the help and support.